\documentclass[letterpaper, conference]{IEEEtran}  
\pdfoutput=1                                                        

\IEEEoverridecommandlockouts                              

\usepackage{graphics} 
\usepackage{graphicx}
\usepackage{subcaption}
\usepackage{array}
\usepackage{epsfig} 
\usepackage{amsmath} 
\usepackage{amssymb}  
\usepackage{color}
\usepackage{float}
\usepackage{algorithm}
\usepackage{algorithmic}
\usepackage{url}
\usepackage{calrsfs}
\usepackage{textcomp}
\usepackage{multirow}
\usepackage{booktabs}
\DeclareMathAlphabet{\pazocal}{OMS}{zplm}{m}{n}
\newcolumntype{P}[1]{>{\centering\arraybackslash}p{#1}}

\DeclareMathOperator*{\argmin}{arg\,min}  

\begin{document}

\title{Is Machine Learning in Power Systems Vulnerable?}

\author{
\IEEEauthorblockN{Yize Chen\IEEEauthorrefmark{1}, Yushi Tan\IEEEauthorrefmark{1}, and Deepjyoti Deka\IEEEauthorrefmark{2}}
\IEEEauthorblockA{\IEEEauthorrefmark{1}Department of Electrical Engineering, University of Washington, Seattle, USA
	\\\{yizechen, ystan\}@uw.edu}
\IEEEauthorblockA{\IEEEauthorrefmark{2}Theory Division, Los Alamos National Laboratory, Los Alamos, USA
	\\deepjyoti@lanl.gov}
}

\maketitle

\begin{abstract}
Recent advances in Machine Learning~(ML) have led to its broad adoption in a series of power system applications, ranging from meter data analytics, renewable/load/price forecasting to grid security assessment. Although these data-driven methods yield state-of-the-art performances in many tasks, the robustness and security of applying such algorithms in modern power grids have not been discussed. In this paper, we attempt to address the issues regarding the security of ML applications in power systems. We first show that most of the current ML algorithms proposed in power systems are vulnerable to adversarial examples, which are maliciously crafted input data. We then adopt and extend a simple yet efficient algorithm for finding subtle perturbations, which could be used for generating adversaries for both categorical~(e.g., user load profile classification) and sequential applications~(e.g., renewables generation forecasting). Case studies on classification of power quality disturbances and forecast of building loads demonstrate the vulnerabilities of current ML algorithms in power networks under our adversarial designs. These vulnerabilities call for design of robust and secure ML algorithms for real world applications. 

\end{abstract}

\section{Introduction}
The modern power systems, with deeper penetration of renewable generation and higher level of demand-side participation, are faced with increasing degree of complexities and uncertainties~\cite{albadi2008summary, patel2005wind}. Reliable operation of the grid in this context calls for improved techniques in system modeling, assessment and decision making~\cite{chen2018model, valtierra2014detection, li2017distributed}. On the one hand, smart meters and advanced sensing technologies have made the collection of fine-grained electricity data, both historical and streaming, available to system operators~\cite{wang2018deep}. On the other hand, there is an urgent need of efficient and near real-time algorithms to analyze and make better use of these available data.

Recent advancements on Machine Learning~(ML) algorithms, especially the giant leaps on deep learning, make ML a good candidate in solving a series of data-driven problems in power systems~\cite{lecun2015deep}. To name a few, ML methods such as Recurrent Neural Networks~(RNN) can find its straightforward applications in wind/solar power and building load forecasting~\cite{hong2016probabilistic, chen2017modeling, chen2017unsupervised}. In~\cite{valtierra2014detection, eskandarpour2017machine}, ML algorithms are applied on power grid outage detection; while in~\cite{wang2018deep}, deep convolutional neural networks are adopted for classifying user load profiles. Planning and control problems in power systems, such as HVAC control and grid protection policy-making, can also be solved via ML approaches~\cite{chen2017modeling, Lassetter2018learning}. All of the algorithms mentioned above have achieved either better performances compared to traditional model-based methods, or have proven to be computationally more efficient. These progresses have shown the great potential of applying ML in power systems.

However, since power systems are at the core of critical infrastructures, we are taking a step back cautiously, and asking ourselves two simple yet not-answered questions:

\emph{``Is ML in power systems vulnerable to data attacks?} 

	\emph{~Are vulnerabilities of ML-integrated power systems easy to deciper by an adversary?" }

\begin{figure}[h]
	\centering
	\includegraphics[scale=0.315]{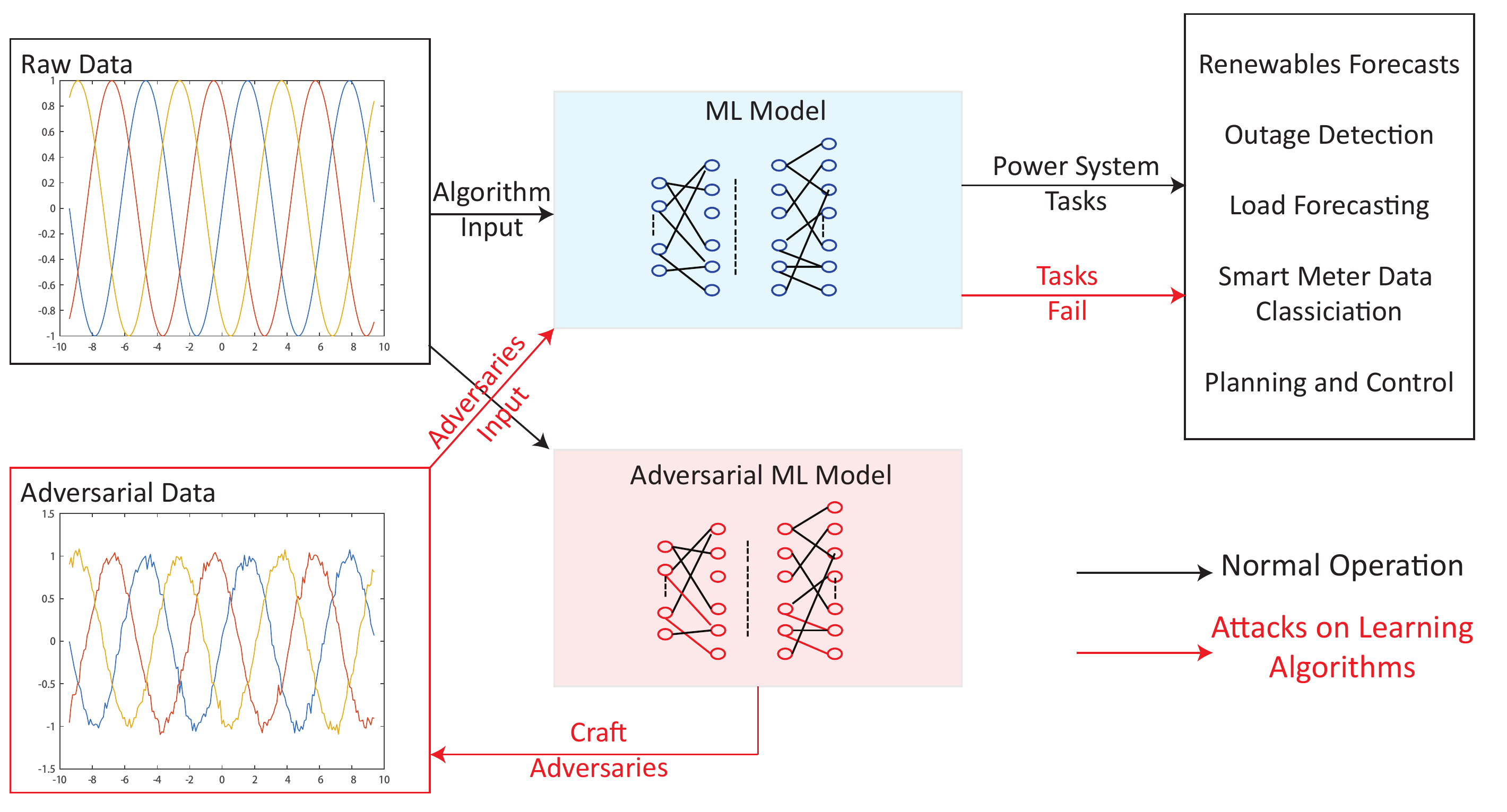}
	\caption{The schematic of the proposed attack on ML in power systems. (Black) Normal ML operations, which learn from the given raw data and has various applications in power systems; (Red) without knowing any knowledge of targeted ML model~(Blue), attackers could generate adversarial examples by only using raw data. Such adversaries would exploit the vulnerabilities of the targeted ML models.}
	\label{fig:intro}
\end{figure}

Unfortunately, in this paper, we answer both these questions affirmatively. By adopting and extending the algorithms proposed in~\cite{papernot2016crafting, hosseini2017blocking}, we show that most of the ML algorithms designed for power systems are vulnerable to adversarial data manipulation, often under very weak assumptions on adversarial ability. As depicted in Fig.~\ref{fig:intro}, attackers do not need any access to the operating ML model itself. Using limited access to the input data, one can generate adversarial data by \emph{injecting designed perturbations} to the original data. The operating ML model's performance~(e.g., classification accuracy) is greatly impaired with these adversarial inputs. 

To demonstrate that such vulnerabilities broadly exist in currently proposed ML algorithms for power systems, we show two typical cases on categorical and time-series cases respectively. In the first case, we successfully attack a power quality disturbances classifier~\cite{valtierra2014detection, eskandarpour2017machine}, which leads to a misclassification of over $70\%$ of given adversarial voltage signals~(e.g., label sage signals as normal). In the second case, we consider an RNN-based building load forecasting model~\cite{chen2017modeling, hahn2009electric}. After imposing crafted perturbations on input variables such as the temperature setpoints and building occupancies, the attack results in a significant performance degradation in the sense that the predicting accuracy drops by a factor of ten. The adversaries in both cases thus exhibit detrimental impacts on power system operations.

\subsection{Contributions}
In the area of computer vision, researchers have found that Neural Network models behave poorly on some crafted images created by simply adding noises to clean images~\cite{goodfellow2014explaining}. This kind of misbehavior on noisy input may be more hazardous for highly automated power systems, since one single wrong decision made by the ML model could undermine the secure operation and lead to a large scale blackout. In light of the criticality of secure sensing and estimation in power grids, this paper includes the following key contributions:
\begin{itemize}
	\item We highlight and discuss the general security issues of ML algorithms in power systems. 
	\item We propose an efficient attacking strategy, which could find the vulnerabilities of ML algorithms in both static and transient cases.
	\item We provide detailed numerical simulations of the proposed adversarial algorithm design, which reveal the vulnerabilities of current ML approaches. We also open-source our code for reproducing the results and testing the security of other physical-ML integrated systems\footnote{Code repository: https://github.com/chennnnnyize/PowerAdversary}.
\end{itemize}

The rest of the paper is organized as follows. In Section~\ref{sec:model} we discuss the general model setup for learning problems in power systems; in Section~\ref{sec:attack} we describe our implementations of attacks on ML models; in Section~\ref{sec:case} we show two representative cases of vulnerabilities on current algorithms; we draw conclusions in Section~\ref{sec:conclusion} with more discussions on the security and robustness of ML model in power systems.

\section{Machine Learning in Power Systems}
\label{sec:model}
In this section, we briefly review the ML models of interest, along with the specific model architecture in the case of Neural Networks. We also introduce the model setup of Recurrent Neural Networks~(RNN), which is a powerful modeling and learning algorithm for sequential data.

\subsection{Learning Task}
Machine Learning provides tools for learning the patterns or relationship in available data, which can be generalized to the future operation and decision-making in power systems. The \emph{supervised learning} setup is normally considered, where a paired training dataset ${X,Y}$ is given. $X,Y$ are vectors of fixed dimensions. For instance, in the case of power quality classification, $X$ are the collected fixed-length power signals, while $Y$ are the one-hot encoded vectors of respective labels~\cite{valtierra2014detection}. The ML model aims to learn a function $f_{\theta}(X)$ that maps from $X$ to $Y$ with model parameters $\theta$. For convenience, we sometimes suppress the $\theta$ symbol. In order to find such mapping, we consider the general algorithm,
\begin{equation}
\label{equ:1}
\theta^*=\argmin_{\theta} \;L(f_{\theta}(X),Y)
\end{equation}
where $L(\cdot,\cdot)$ is a pre-defined loss function. For instance, $L_2$ loss can be directly used to determine the distance, which is commonly used in LASSO along with $L_1$ regularization on $\theta$; while in the case of classification using Neural Networks, we may choose cross-entropy for $L(\cdot,\cdot)$, and in the case of regression via Neural Networks, an $L_2$ loss is feasible to determine the deviation of model's outputs from true values. 

Since many of the ML applications~\cite{Lassetter2018learning, chen2017modeling, valtierra2014detection} have focused on utilizing the learning and representation capabilities provided by Neural Networks, here we briefly illustrate the learning procedure on Neural Networks' parameters~(neurons' weights). Neural Networks are composed of stacked, differentiable ``neuronal" layers, such as fully connected layers, convolutional layers and activation functions. It is powerful in learning tasks with high-dimensional $X$ and $Y$. Though there are many variants of the iterative steps, the standard back-propagation procedure via gradient descent for updating model weights is summarized as follows,
\begin{equation}
\label{equ:2}
\theta_{i+1}=\theta_{i}-\eta \nabla_{\theta} L(f_{\theta}(X),Y)
\end{equation}
where $\eta$ is the learning rate, and the subscripts on $\theta$ denote the iteration steps on the weight parameters of the Neural Networks. Once the model is trained using $X,Y$ via \eqref{equ:1} and \eqref{equ:2}, we get an accurate model $f_{\theta^*}$. Recent progresses on deep learning have enabled Neural Networks composed of millions of neurons to outperform all other algorithms in many real-world applications~\cite{lecun2015deep}.

\subsection{RNN Model}
\begin{figure}[h]
	\centering
	\includegraphics[scale=0.51]{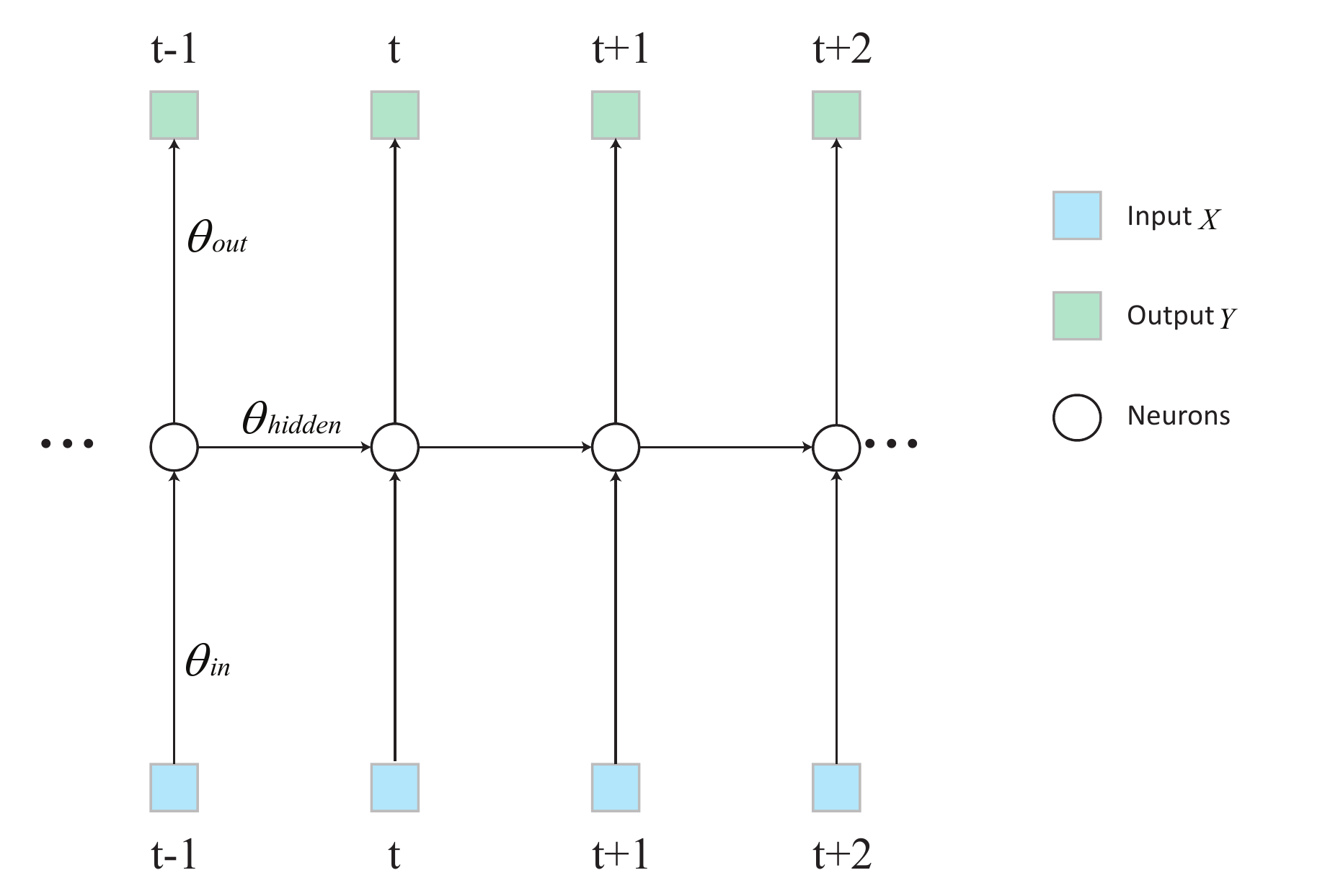}
	\caption{Basic RNN structure composed of hidden, input and output neurons. The output $\hat{y}_T$ is a function of sequential input $x_0,...,x_T$. with a memory length $T$.}
	\label{fig:rnn}
\end{figure}

In many cases, the states in power systems are not static, but rather evolve in a sequential manner. For instance, the future solar power and wind generation have temporal and spatial correlations. Under this scenario, Recurrent Neural Network~(RNN) becomes a good fit as its structure allows it to model temporal dependencies for sequence data~\cite{mikolov2010recurrent}. 

Modeling via RNN requires a group of sequential input sample $x=\{x_0,...,x_T\}$, where $T$ is the memory length. The weight coefficient of RNN consists of three subsets: $\theta_{in,t}, \theta_{out,t}$ and $\theta_{hiddent,t}$. RNN also allows for linked neurons between neighboring timesteps. The t-timestep RNN cell is using inputs from hidden state $h_t$ and input $x_t$, and delivers outputs $\hat{y}_t$ as well as next step's hidden state $h_{t+1}$. The $t$-step RNN cell then completes the following computations:
\begin{subequations}
	\begin{align}
		\hat{y}_t&=f_{\theta_{in,t},\theta_{out,t}}(x_t,h_t)\\
	h_{t+1}&=f_{\theta_{hidden,t},\theta_{in,t}}(x_t,h_t)
	\end{align}
\end{subequations}

By stacking such cells over time, the hidden state can be used to store and transfer the input information from previous steps. With a memory length of $T$, the output $\hat{y}_T$ is essentially a function of $x_0,...,x_T$. We can then conclude that RNN's modeling and learning strategies also take the form of~\eqref{equ:1} and~\eqref{equ:2}, where $X$ is composed of the sequential vectors $\{x_t,\,t=0,...,T\}$.

\section{Crafting Attacks for ML}
\label{sec:attack}
In this section, we first give mathematical definitions on \emph{adversarial examples} which exploit ML's vulnerabilities. We then propose an algorithm, which is a variant of the Neural Networks attack approach proposed in~\cite{goodfellow2014explaining}. Our proposed algorithm produces adversarial examples for both normal Neural Networks and sequential models such as RNN.  

\subsection{Adversarial Examples}
Consider any given supervised ML model $f_{\theta}$ with corresponding paired dataset $X,Y$. We assume that an attacker has \emph{no} access to the model $f$ and cannot modify it. Instead we consider the mild setting that the attacker can only change the input samples $X$ to $X^*$ yt{to the model to modify its output $f_{\theta}(X^*)$ such that is not accurate compared to the ground truth $Y$}. Moreover, to avoid detection by the system operator, the attacker ensures that adversarial input $X^*$ is close to the true inputs $X$. For instance, an attacker tries to modify the system voltage wavelet signals such that ML-based power quality classifier would classify falsely, while making sure that such changes on signals would not be observed by the system operator. Formally, the attacker would craft an adversary via solving the following optimization problem:
\begin{subequations}
	\label{equ:4}
	\begin{align}
	\max_{\delta_X} \;\;\;&L(f_{\theta}(X^*),Y)\\
	s.t. \;\;\;&X^*=X+\delta_X \label{equ:defAdver}\\
	&||\delta_X||_d\leq \gamma\cdot |X| \label{equ:errors}
	\end{align}
\end{subequations}
where $\delta_X$ in~\eqref{equ:defAdver} is the perturbation we add to the clean samples $X$; \eqref{equ:errors} constrains the level of perturbation $\gamma$ allowed for $X^*$. Different choices of $d$ for the norm of $\delta_X$ lead to different constraints on adversarial manipulation:
\begin{itemize}
	\item $d=0:$ \eqref{equ:4} has similar objective as the Grad$_0$ attack proposed by~\cite{hosseini2017blocking}, where $\gamma$ denotes how many dimensions of input data is allowed to be modified.
	\item $d=\infty:$ \eqref{equ:4} has similar objective as the Fast Gradient Sign~(FGS) attack proposed by~\cite{goodfellow2014explaining}, where $\gamma$ denotes the maximum level of noise allowed on each dimension of $\delta_X$.
\end{itemize} 

We also observe an interesting connection between \eqref{equ:2} (operator) and \eqref{equ:4} (adversary), where the ML training algorithm is essentially training over model parameters $\theta$ to minimize model loss; while the adversaries' task is quite opposite: to optimize over model \emph{inputs} $X$ to \emph{maximize} model loss. Specifically, we look into the case of Neural Networks involving highly non-convex model in terms of both $X$ and $\theta$ that have been shown to achieve state of the art performance in several power system applications. Since solving \eqref{equ:2} always yields an accurate model, we are interested in finding ways to solve \eqref{equ:4} which would provide insights on the vulnerabilities of Neural Networks used in power systems.

\subsection{Crafting Adversarial Examples}
\label{attacker}

In this sub-section, we propose an efficient attack algorithm which can incorporate the constraints~\eqref{equ:errors} with $d=0$ and $d=\infty$ and exploit the vulnerabilities for both normal Neural Networks and sequential models like RNN.

\begin{figure*}[!t]
	\centering
	\includegraphics[scale=0.75]{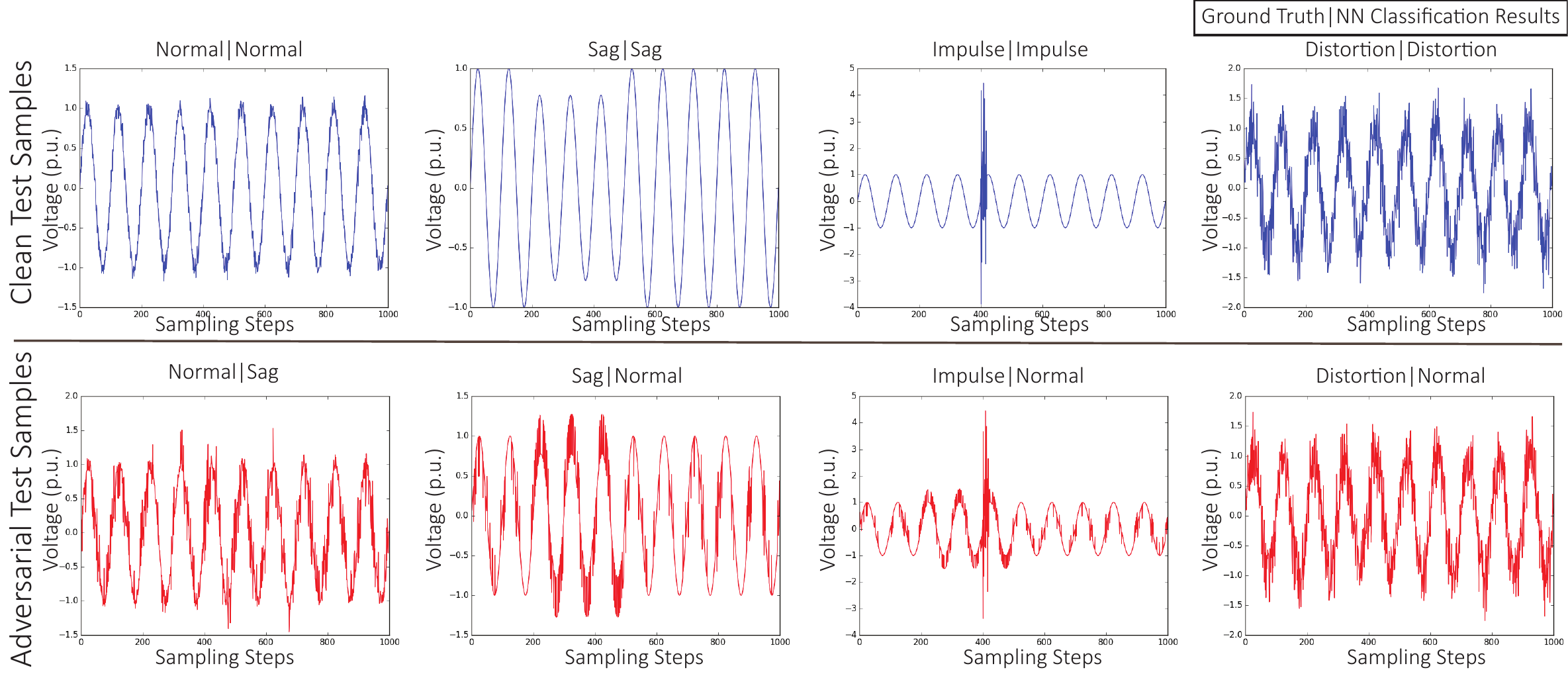}
	\caption{\small
		{
			Case studies on power quality signal classification with randomly selected clean samples from our test sets (top) versus corresponding adversarial samples crafted by Algorithm.~\ref{algo} (bottom). The original Neural Networks could accurately classify four classes of power signals, yet it fails to classify adversarial samples with high probability.}
	}
	\label{fig:signal}
\end{figure*}

\subsubsection{Adversarial Samples without $d=0$  Constraint}
Since the optimization problem \eqref{equ:4} itself is highly nonconvex and high-dimensional, it is intractable to achieve the global optimal solution $X^*$. Alternatively, since the gradients of $L(f_{\theta}(X),Y)$ encode the loss landscape, we propose a gradient ascent method on the loss function with respect to $X$ to acquire the small perturbations which would increase $L(f_{\theta}(X),Y)$:
\begin{equation}
\label{equ:5}
X^*=X+\delta_X=X+\epsilon \nabla_X L(f_{\theta}(X),Y)
\end{equation} 
where $\epsilon$ controls the noise level added to the clean samples. Crafting attacks following~\eqref{equ:5} exactly follows the FGS attack strategy, which has found vulnerabilities for ML models used in computer vision. Yet this attack has no constraint on $||\delta_X||_0\leq \gamma\cdot |X|$, so attacker has the control and access to modify every entry of $X$, which adds relatively large perturbations to the input.

\subsubsection{Adversarial Samples with $d=0$ Constraint}
We now discuss the constraint on the number of entries the attacker is allowed to modify. The attacker shall only change the $\gamma\cdot|X|$ input entries, which have the most impact on $L(f_\theta(X+\delta_X), Y)$. Formally, let $A$ define the set of largest $\gamma \cdot |X|$ entries of $\nabla_X L(f_\theta(X+\delta_X), Y)$, while let $S$ denote the entire set of entries. Then we propose the following operation to get adversarial samples with constraint $||\delta_X||_0\leq \gamma\cdot |X|$:
\begin{subequations}
	\begin{align}
	&\delta_X^{A}=\epsilon \nabla_{X^A} L(f_{\theta}(X),Y)\\
	&\delta_X^{S \backslash A}=0
	\end{align}
\end{subequations}
$S\backslash A$ denotes the complement set of input entries. The final adversarial examples can still be generated via $X^*=X+\delta_X$. Since all the ML models considered in this paper, including normal Neural Networks and RNN structures are all differentiable with respect to input, we highlight the universality of the proposed algorithm on finding the vulnerabilities of any trained models. 

Note that even though \eqref{equ:5} only implements once in our proposed algorithm without any iterative optimization on $\delta_X$, we show in Section~\ref{sec:case} that the trained, unknown model is vulnerable to such attacks. 

We also distinguish our work from previous attack and defense research in power systems~\cite{liu2011false, huang2013bad}. Previous research only exploited the vulnerabilities of state estimation, while we found weaknesses of general ML tasks in power systems. Moreover, the proposed algorithm works under the \emph{black-box} setting. To put it in other words, the attacker only needs to train its own version of surrogate ML model $f_{\theta}'$ without knowing any knowledge of  $f_{\theta}$. By finding adversarial examples $X^*$ of $f_{\theta}'$, $X^*$ can then be used for attacking unknown ML model $f_{\theta}$ operating in power systems. We summarize the algorithm in Algorithm~\ref{algo}.

 \begin{algorithm}
	\caption{Crafting Adversarial Examples}
	\label{algo}
	\begin{algorithmic}
		\REQUIRE Clean pairing training data $X,Y$, input entries set $A$
		\REQUIRE Training iterations $N_{iter}$, number of adversarial examples $N_{adv}$
		\REQUIRE Clean testing samples $\{x_i,y_i\}, i=1,...,N_{adv}$
		\ENSURE Attacker surrogate ML model $f_{\theta}'$
		\ENSURE Adversarial examples set  $X^*\leftarrow \emptyset$
		\STATE \# Training the surrogate ML model
		\FOR {iteration$=0,...,N_{iter}$}
		\STATE Update $\theta'$ using gradient descent \eqref{equ:2} on $X,Y$
		\ENDFOR
		\STATE \# Find adversarial examples using clean data $\{x_i,y_i\}$
		\FOR {iteration $=0,...,N_{adv}$}
		\STATE Calculate gradients w.r.t $x_i$: $\delta_{x_i}=\epsilon \nabla_{x_i} L(f_{\theta'}(x_i),y_i)$
		\STATE Find set $A$: the largest $\gamma\cdot |X|$ gradients of $\delta_{x_i}$
		\STATE {$\delta_{x_i}^{S \backslash A}=0$}
		\STATE {$x_i^*=x_i+\delta_{x_i}$}
		\ENDFOR
		
		\STATE $X^*$.insert($x_{i}^*$)
	\end{algorithmic}
\end{algorithm}

\section{Case Studies}
\label{sec:case}
We evaluate the proposed algorithm's performance on two tasks: power quality assessment via classifying the voltage signals by feed-forward Neural Network~\cite{valtierra2014detection, eskandarpour2017machine}, and short-term building load forecasting via RNN~\cite{chen2017modeling}. We set up the deep learning models using Tensorflow and Keras, two Python open-source packages. We adopt rectified linear unit~(ReLU) activation functions, dropout layers and Stochastic Gradient Descent, a variant of \eqref{equ:2} to improve the performance of our Neural Networks model. Two Nvidia Geforce GTX TITAN X GPUs are used for training acceleration and the average training times of both tasks are within $10$ seconds.

\subsection{Power Quality Classification}
In this task, we would like to investigate if ML model could detect the power quality disturbances in the waveform signals. Past research claim that Neural Networks based classifier would detect those disturbances in signals, which would then avoid damages and improve the power quality~\cite{valtierra2014detection, eskandarpour2017machine}. Here, we attempt to add slight perturbations to the input signals and see if such classifier would fail to classify these disturbances.

\subsubsection{Data Description}
\begin{figure}[h]
	\centering
	\includegraphics[scale=0.49]{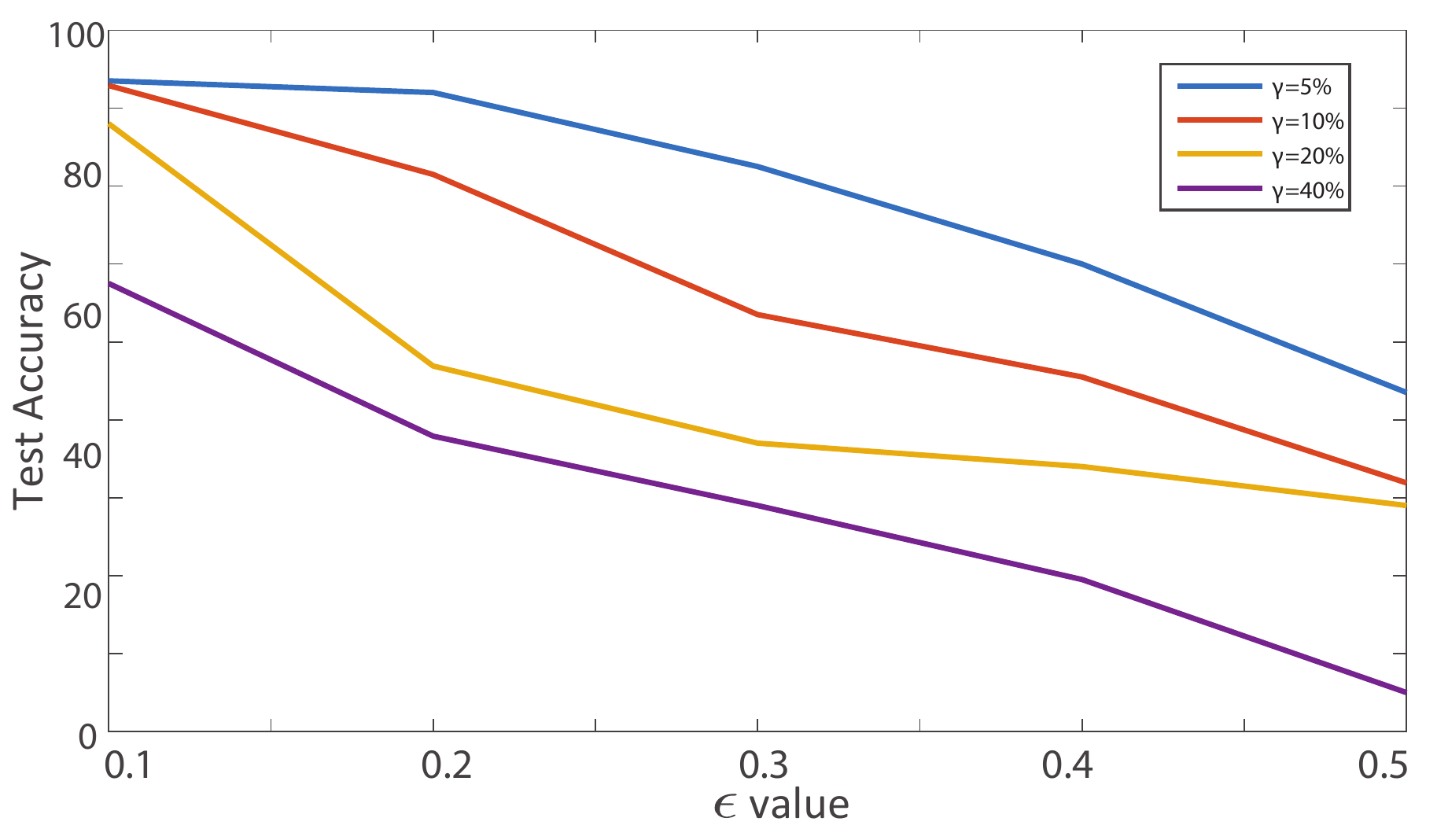}
	\caption{Voltage signal classification accuracy with varying noise level $\epsilon$ and input perturbation percentage $\gamma$ for adversaries $X^*$.}
	\label{fig:reg_result}
\end{figure}

We consider four types of wave signals as illustrated in the first row of Fig.~\ref{fig:signal}, with one group of normal signals, and three types of disturbances: sags, impulses and distortion. We construct a labeled dataset with $200$ signals from each class, with each signal of fixed length. After shuffling and separating $\frac{1}{4}$ of the data as testing set, we construct a $3-$layer fully connected Neural Networks to classify these signals into their respective class.

\subsubsection{Simulation Results}
We firstly observe the Neural Networks classifier is powerful in classifying wave signals with different source of disturbances. The model achieves $97.5\%$ testing accuracy on the split test data. 

Then we test if such trained classifier is able to correctly classify the adversarial signals crafted by Algorithm~\ref{algo}. As shown in Fig.~\ref{fig:signal}, with $\epsilon=0.03$ and $\gamma=10\%$, the black-box classifier wrongly classifies the adversarial signals. Specifically, the adversarial impulse and distortion signals look similar to corresponding clean signals, and can be still classified as impulse and distortion signals by a technician, yet the ML model incorrectly regards them as normal signals. 

As shown in Fig.~\ref{fig:reg_result}, we qualitatively test the adversaries' performance by evaluating the Neural Networks' classification result on adversarial examples. The model's classification accuracy drops drastically with higher level of $\epsilon$ and $\gamma$, which meets our assumption. When $\gamma=40\%$ in which our algorithm changes $40\%$ entries of the input signal, by only injecting a small perturbation $\epsilon=0.1$, the ML model can only classify $67.5\%$ of the test samples.

\subsection{Building Load Forecasting}
In this example, we first train an RNN model, which could forecast building load accurately by using input features such as temperature measurements, building occupancy and solar radiation. We then construct sequential adversarial inputs by using a surrogate model and evaluate the vulnerabilities on the original load forecasting model.

\begin{figure}[h]
	\centering
	\includegraphics[scale=0.31]{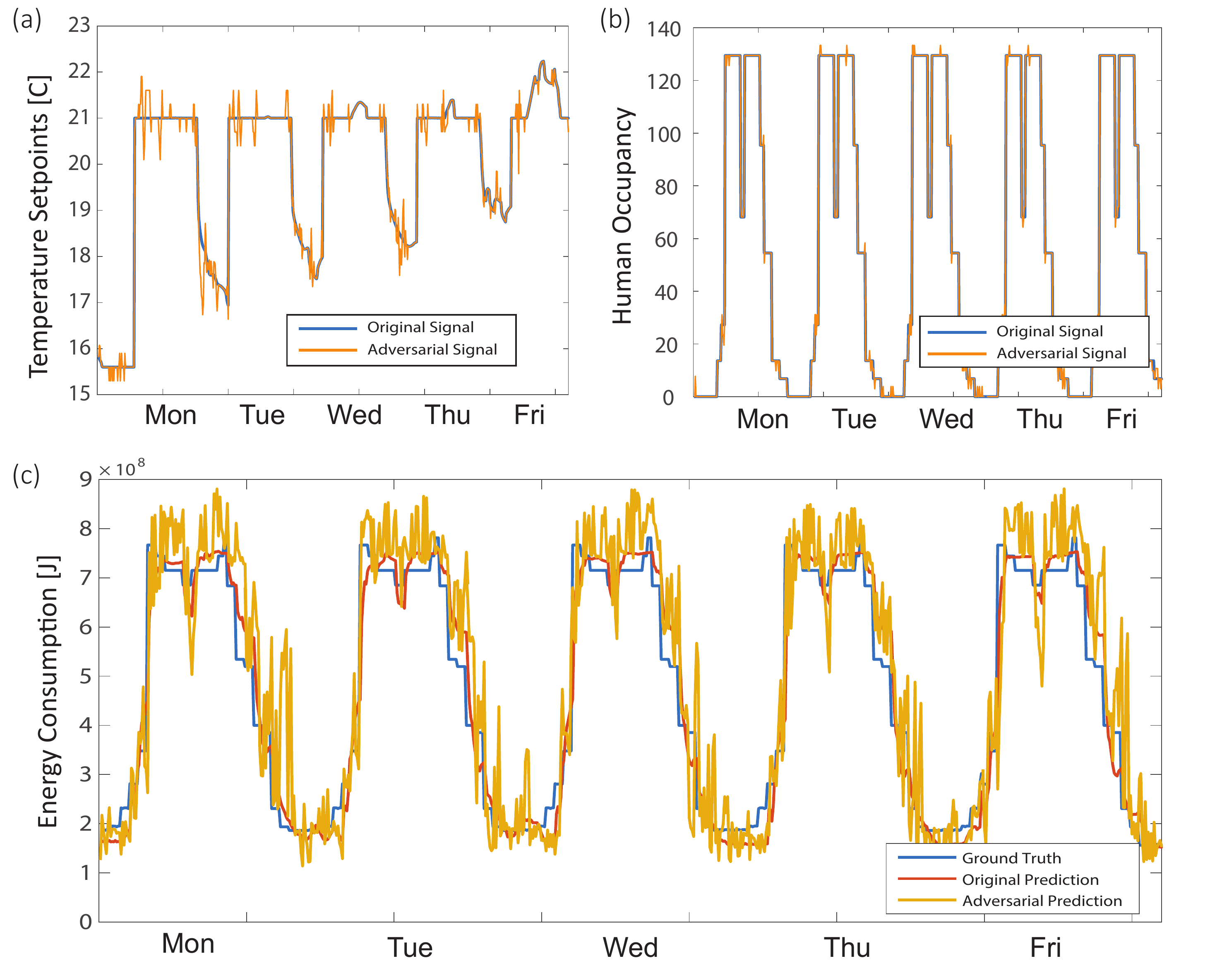}
	\caption{Building forecasts results under $\epsilon=0.03$ and $\gamma=10$. (a) and (b) the data profiles for one week's sub-region temperature setpoints and occupancy level before and after the attack; (c) the ground truth of one week's energy consumption, predicted energy consumption using clean testing data and predicted result after injecting adversarial data profiles.}
	\label{fig:building}
\end{figure}

\subsubsection{Data Description}
We set up our building simulation platforms using EnergyPlus's 12-storey large office building listed in the commercial reference buildings from U.S. Department of Energy (DoE CRB)~\cite{torcellini2008doe}. The building has a total floor area of $498,584$ square feet which is divided into $16$ separate zones. We simulate the building running through the year of $2004$ in Seattle, WA, and record $x_t$, $y_t$ with a resolution of $10$ minutes, where $x_t$ includes data coming from various sensors, such as building occupancy, temperature setpoint and temperature measurements, and $y_t$ is the building energy consumption.  We shuffle and separate $2$ months' data as our stand-alone testing dataset for both predictive accuracy validation and vulnerabilities testing. 
The RNN model is composed of $1$ recurrent layer and $2$ subsequent fully-connected layers with a memory length of $2$ hours. Our ML model is also easy to extend to Long Short-Term Memory network~(LSTM) or any other variants of RNN structure. Since all these architectures are differentiable w.r.t $x_t$, they would exhibit similar vulnerabilities to proposed adversaries.

\vspace{-10pt}
\begin{table}[!htbp]
	\centering
	\begin{tabular}
		{P{1.2cm}| P{1.5cm}|P{1.5cm}|P{1.5cm} }
		\hline
		\toprule[0.5mm]
		MAPE &Temp Deviation  & Occupancy Deviation   & Prediction Error \\
		\hline 
		$\epsilon=0.0$ & $0\%$ & $0\%$ & $5.29\% $ \\
		$\epsilon=0.01$ & $0.35\%$ & $2.44\%$ & $25.90\% $ \\
		$\epsilon=0.03$ & $1.07\%$ & $6.94\%$ & $31.55\% $ \\
		$\epsilon=0.05$ & $1.86\%$ & $12.36\%$ & $55.37\% $ \\
		
		\toprule[0.5mm]
	\end{tabular}
	\caption{The building load forecasting performance using adversarial examples with varying noise level $\epsilon$ under $\gamma=10\%$. Note when $\epsilon=0$ it is the case with clean testing data.\vspace{-10pt}}
	\label{table}
\end{table}

\subsubsection{Simulation Results}
We use the \emph{Mean Absolute Percentage Error}~(MAPE) to evaluate both the forecasting error and the input feature deviation caused by adding adversarial perturbations:
\begin{equation}
MAPE(var^*,var)=\frac{1}{N} \sum_{i=1}^{N} \frac{|var^* - var|}{var}\times 100\%
\end{equation}
where $var$ represents either input feature or output energy consumption, while $var^*$ represents the corresponding adversarial feature or output energy consumption prediction. We test the ML model performance by using the same one-week of testing data with different level of $\epsilon$ on adversarial data.

As can be seen in Table.~\ref{table}, the model performs well with only $5.29\%$ MAPE by using clean data. However, by only injecting $\delta_X$ with $\epsilon=0.01$, the model's forecast has a $25.90\%$ deviation from the ground truth. The results worsen with more intense level of noise injected. Meanwhile, the input features have little deviation from the clean data. We can also visually inspect the vulnerabilities of the RNN model in Fig.~\ref{fig:building} when we only change $10\%$ of the input features with noise level $\epsilon=0.03$. The output prediction jumps a lot compared to previous forecasts, which is not informative for building operators.

\section{Conclusion and Discussion}
\label{sec:conclusion}
In this work, we look into the security and vulnerability of Machine Learning algorithms in power systems to adversaries. We propose an attack algorithm that universally exploits the vulnerabilities of ML in power systems, especially Neural Network based algorithms. The adversarial strategy is practical as it does not change the system operator's  ML engine but manipulates only input data. Case studies on two representative power system examples reveal the vulnerability of proposed ML algorithms. As researchers haven't looked into such vulnerabilities in current algorithm design, we hope our work will stimulate future discussions to increase the robustness of current ML algorithms in power systems to data manipulation. Going forward, the following directions regarding secure ML applications are worth investigating, and we are also interested in investigating security issues of broader estimation/learning algorithms in power system operation and control.

\subsection{Adversaries in Learning}In this work, we only discuss the scenario that ML models in power systems output inaccurate results with adversarial inputs. Stronger attacks such as \emph{targeted attack} can be considered where instead of solely maximizing the predicted loss, the attacker can also add perturbations to falsify trained models to a new adversarial objective. Such attacks are discussed in our previous work~\cite{hosseini2017blocking} which can be extended to the case in power systems. Moreover, there are also vulnerabilities to the model itself. For instance, attacker would hack into the operation room to change the weights of trained model. Even though there is a line of work addressing the security issues in the control, communication and infrastructure of power systems, there is scope for work to address the security of learning in power and cyber-physical systems. 

\subsection{Defense for ML Algorithms in Power Systems} Up to now, there has been some work on defending ML attacks in the research of computer vision. Yet most of them operate on the ensemble or filtering of input images~\cite{tramer2017ensemble}, which may not be applicable for power systems as most of applications involved in power have clear physical definitions on the features involved that may no be modifiable. Thus the defense against such intrusion attacks on ML algorithms in power systems is still a urgent yet open problem.


\bibliographystyle{IEEEtran}
\bibliography{bib}

\end{document}